**Engineering Sustainability**

# Life cycle assessment tools for road design: analysing linearity assumptions

**Nikolaos Kalyviotis** Dipl Ing, MSc, MSPM, MBA, PhD
Civil Engineer, Technical Studies Department, Directorate of Technical Works,
University of Crete, Rethymno, Crete, Greece (nkalyviotis@uoc.gr)

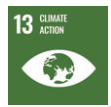

Road infrastructure significantly impacts how people move and live and the emissions associated with travel behaviour. The design of roads is crucial in mitigating emissions. This paper reviews existing transport life cycle assessment tools that have been developed by various entities and can be used for roads. The review focuses on data sources used in the analysis, methods of estimating carbon dioxide emissions, the underlying software that is used to make the estimates, and any limitations of the tools. A critical issue identified in life cycle assessment analysis is the erroneous assumption that relationships within the assessed systems are linear. The current tools focusing on transport infrastructure assessment were developed based on the linear assumptions and limitations of the life cycle assessment analysis. A significant research gap identified is that existing life cycle assessment tools are not integrated with the design process. The analysis is an add-on process to design and the results of an assessment are not then used iteratively to enhance the design. A case study on aggregate road design found that road area significantly correlates with emissions, slope adjustments reduce emissions, and soil type impacts emissions, suggesting future research should explore non-linear relationships for sustainable road design.

## Notation

| | |
|---|---|
| CBR | California bearing ratio |
| $E_{BS}$ | base course resilient modulus |
| $h_{bs}$ | base layer thickness |
| $k$ | modulus of subgrade reaction |
| $M_R$ | resilient modulus |

## 1. Introduction

Embarking on a journey towards a sustainable future, the EU and the UK have set ambitious goals to become climate-neutral by the year 2050 (Borning *et al*., 2020). Central to this mission is the transformation of transport infrastructure—roads, highways, drainage systems, and earthworks—which currently stands as a significant source of embodied emissions (Pantelidou *et al*., 2012).

The UK Net Zero Highways (National Highways, 2021) strategy targets net-zero emissions in maintenance and construction by 2040. This plan emphasises the need to use greener construction materials to create a sustainable road network. As the UK expands its infrastructure, it must balance progress with environmental stewardship, minimising the emissions footprint of new projects. Understanding the current emissions baseline is crucial for aligning road development with climate goals, ensuring each new mile supports a sustainable future.

Life cycle assessment (LCA) is defined as a systematic process for evaluating the environmental impacts associated with all the stages of a product's life cycle (ISO, 1997). This includes everything from raw material extraction (the 'cradle') through manufacturing, distribution, and use, to the end-of-life disposal or recycling (the 'grave') (ISO, 1997). The methodology involves a detailed inventory of energy and material inputs and environmental releases, assessing the cumulative potential environmental impacts. LCA methods vary in their data requirements, uncertainty, and assumptions (Kalyviotis, 2022).

LCA began in the 1970s to quantify the environmental impact of products through pollutants. By the 1990s, it evolved to include indirect factors such as energy use (Kikuchi, 2016). In 1995, Society of Environmental Toxicology and Chemistry (SETAC) defined LCA as evaluating environmental burdens by identifying and describing energy and materials used, and wastes released, throughout a product's life cycle, from raw material extraction to final disposal (Kikuchi, 2016). This includes manufacturing, distribution, use, reuse, maintenance, recycling, and transportation. LCA addresses impacts on ecological systems, human health, and resource depletion. The concept is often summarised as 'cradle-to-grave' (Curran, 1997; Curran and Young, 1996).

The International Organization for Standardization (ISO) created an LCA standard in 1997, revised in 2006 (ISO, 2006). According

to ISO (2006), LCA addresses environmental aspects and potential impacts of a product by inventorying input/output data, providing additional information to assess and understand the environmental significance and summarising results for conclusions, recommendations, and decision making.

The 2006 revision expanded the focus to include environmental aspects and impacts throughout a product's life cycle, from raw material acquisition to final disposal ('cradle-to-grave'), aligning with SETAC's definition of LCA. The challenge is applying this to complex systems such as infrastructure.

Transport infrastructure can be comprehensively defined by synthesising the insights from various scholars and institutions: transport infrastructure represents the foundational public works essential for societal functioning, encompassing roads, bridges, dams, mass transit systems, sewage, and water systems (Jessen, 1984). It is a critical provider of basic services to industries and households, facilitating the daily operations of a modern economy (Martini and Lee, 1996). Characterised as complex, large-scale, interconnected, open, sociotechnical systems, transport infrastructure is a large-scale physical resource constructed by humans for public consumption (Frischmann, 2012). Ultimately, transport infrastructure is integral to the planet's overall functioning, supporting the movement and connectivity that drive global progress and well-being (Allenby and Chester, 2018). This definition encapsulates the multi-faceted nature of transport infrastructure, highlighting its role as a pivotal element in the tapestry of human civilisation.

A common way to compute the emissions generated is the usage of rational multipliers (Merciai and Schmidt, 2018, 2016). The selection and estimation of appropriate multipliers are crucial for accurately determining the emissions generated by each sector or activity (Merciai and Schmidt, 2018; Stadler et al., 2018). This approach has partly been adopted from the LCA approach using characterisation factors. Characterisation starts with the categorisation of the environmental impact in the proper category, followed by the application of the corresponding model to indicate the results of the category (ISO, 2006). In other words, characterisation is the calculation of category indicator results by summing the products of characterisation factor times the inventory results of each category. Then, normalisation of the results is achieved by presenting them to a relative reference value over a defined period of time (ISO, 2006). After the normalisation, the results may be weighted by modifying the results of each category to aggregate results using fixed multipliers between the different impact categories or indicators.

This is the logic behind LCA and on this logic were based all the methodologies developed. Even the latest standards, such as the PAS 2080 standard for carbon management in infrastructure (The Green Construction Board, 2023), are based on these principles.

The key element of the LCA methodologies is the conversion of the complicated emission generation relations to linear relations since it regards all processes as linear (Guinée et al., 2001). All the methods use characterisation factors, which 'linearly express the contribution of a unit mass of an emission to the environment' (Menoufi, 2011). Some steps of LCA methodologies that use non-linear equations can still be transformed into linear ones. One common way of quantification met in the LCA methods or tools is the multiplication of two or more parameters or factors. This quantification can be easily transformed to linear by taking the sum of the logarithm functions of the parameters. However, some elements cannot be transformed to linear relations such as the non-linear functions of 'the intermittent character of rainfall or the differences between indoor and outdoor emissions' (Jolliet et al., 2003), potentially affected fraction of species in ECO-INDICATOR 99 method, fuel consumption and emissions profiles with vehicle speed (Chester, 2008; Chester et al., 2010), economic mechanisms (Guinee et al., 2011), and ecosystem restoration times (Guinee et al., 2011).

This paper aims to review existing LCA tools used in road design, with a particular focus on their linearity assumptions. The paper has the following objectives:

- Review of LCA tools: conduct a comprehensive review of current LCA tools utilised in road design.
- Identification of linearity issues: highlight and analyse the issues related to linearity assumptions within these tools.
- Statistical analysis: perform a statistical analysis to examine the linear relationship between various factors and the emissions generated; using a case study of designing aggregate roads based on the American Association of State Highway and Transportation Officials (AASHTO) guidelines.
- Recommendations for new tool development: suggest opportunities and factors to be considered in the development of a new LCA tool that addresses the identified linearity issues.

## 2. Existing transport infrastructure LCA tools

The LCA methods has the issue of the 'illusion of linearity', meaning the tendency to, wrongly, imply linear relationships and apply their properties (Kalyviotis, 2022). The phenomenon may also be reported in the literature as 'linearity trap' or 'linear obstacle'. Freudenthal (1983: p. 267) noticed that 'linearity is such a suggestive property of relations that one readily yields to the seduction to deal with each numerical relation as though it were linear'. The current tools focusing on transport infrastructure assessment were developed based on the assumptions and the limitations of the, aforementioned, LCA methods.

'At heart, LCA is a tool based on linear modelling' (Guinée, 2001). The central aspect of LCA methodologies is the simplification of complex emission generation relationships into linear models, as they treat all processes as linear (Guinée, 2001). These methods employ characterisation factors that linearly represent the environmental impact of a unit mass of an emission (Menoufi, 2011). Even though some steps in LCA methodologies involve non-linear equations, they can often be converted into linear forms. A typical approach to quantification in LCA methods or tools involves multiplying two or more parameters or factors. While other non-linear relationships can be linearised, certain elements resist such simplification. For example, the non-linear functions representing the sporadic nature of rainfall or the variance between indoor and outdoor emissions in the IMPACT 2002 + method (Jolliet *et al*., 2003), or the relationship between fuel consumption, emissions profiles, and vehicle speed (Chester, 2008; Chester *et al*., 2010), as well as economic mechanisms and ecosystem restoration times, cannot be linearised. The Centrum voor Milieukunde Leiden-Impact Assessment (CML-IA) method acknowledges the limitations of linear relationships (Guinée, 2001), which validates the use of characterisation factors (multipliers) to quantify a process's impact (Guinée, 2001; Guinee *et al*., 2011). Furthermore, normalisation and weighting are also applied linearly by multiplying the designation coefficients with the corresponding emissions (Guinée, 2001). The linear nature of the LCA methods is evident, with normalisation occurring per resource unit.

It has been observed that the assumption of linearity is often made when data are insufficient or when the data set is large and detailed, posing challenges in creating an explicit model. In transport infrastructure, the linearity assumption is applied to material usage against emissions based on spatial placement (volume, distance, area), such as for concrete, steel, herbicides, and salt, and for smaller infrastructure projects such as track construction, lighting, and parking spaces (Chester, 2008). It is within this context that the linearity assumption is frequently applied in transport infrastructure assessments.

Most LCA models for transport infrastructure are normalised by distance, using functional units such as passenger-kilometre/mile travelled, track kilometre/mile travelled, or construction length. Even studies that consider various types of infrastructure, such as the research by National Rail (2009), present results normalised by distance, thereby extending the linearity assumption to normalisation. A significant challenge arises when comparing different types of transport infrastructures that can yield biased outcomes.

Transportation LCA tools are specifically designed to enable the practical calculation of the embodied emissions associated with transport infrastructure and include built-in assumptions based on the experience and training datasets of the developers (Hoxha *et al*., 2020).

A comprehensive review of scholarly literature has identified various LCA tools currently used to evaluate transport infrastructure (Kalyviotis and Saxe, 2020). Only tools relevant to road design were retained, and the selection was updated to incorporate new tools and the latest versions of existing ones. Table 1 summarises these tools, detailing their development locations and entry formats. Most tools originate from North America and Europe. The author evaluated each LCA tool's efficacy in different working environments, examining their characteristics. Most tools are spreadsheets that use environmental indicators (e.g. emission indicators) to calculate infrastructure emissions based on the energy consumption of materials, equipment, services, and design. Spreadsheets are common because the analysis relies mainly on linear algebra, which can be easily represented in this format.

The background database for each tool is often specific to that tool. For instance, Atkins' Carbon Critical Knowledgebase uses a proprietary database developed by Atkins. Notably, UK tools were cross-developed using elements and databases from existing tools. Databases such as the Bath Inventory of Carbon and Energy, AggRegain Carbon Dioxide ($CO_2$) Emissions, the British Department for Environment, Food and Rural Affairs (Defra) database, and the British Cement Association database were developed by different organisations but are used by new tools. This approach reduced development costs by leveraging existing background databases.

Among the nine summarised tools, seven encompass the entire cradle-to-grave life cycle stages, including material extraction, construction, operation, maintenance, and end-of-life phases (see Table 1). Two tools focus solely on the cradle-to-construction stages. These tools, which estimate emissions up to the construction phase, were developed either by the private sector or organisations with a vested interest in construction. Two tools calculate emissions for both vehicles and infrastructure, though both rely on data with high uncertainty. The Transportation LCA tool is based on academic data specific to California, while VICE 2.0 utilises broad empirical data.

The inputs required for using the transit LCA tools can be grouped into three categories: (*1*) quantity of materials, (*2*) equipment details and energy consumption (e.g. operating time), and (*3*) infrastructure project design (e.g. project length for the Federal Highway Administration (FHWA) Infrastructure Carbon Estimator). These inputs are estimated at the project's planning stage, and any discrepancies between planned and actual values are reported through the project control process. In essence, all tools consider the energy and materials needed to construct various structural elements, with some also considering the design of these elements. The accuracy of data collection directly influences the precision of the embodied emissions assessment.

This analysis highlights the need for more comprehensive tools capable of analysing road infrastructure, a gap this paper aims to address.

## 3. Methodology

This research uses aggregate-surface roads, since the researcher designed or involved in the design of more than 200 aggregate-surface roads around the world, including all the accompanied data such as geotechnical studies, flood risk assessment studies, environmental studies, and topographical data. The aggregate road design against the asphalt road design has only one 'design step' difference

Table 1. Transportation LCA tools location, entry format and database, boundaries, type, inputs, and description and characteristics (adapted from Kalyviotis and Saxe, 2020)

| Name of tool | Location | Entry format | Background databases | Life cycle stages | Type of tool | Inputs | Description and characteristics |
|---|---|---|---|---|---|---|---|
| asPECT (Asphalt Pavement Embodied Carbon Tool) (Transport Research Laboratory (TRL), 2014) | UK | Spreadsheet | asPECT's database (developed for the tool) | Cradle-to-grave | Tool based on material quantities | Material quantities | The asPECT (Asphalt Pavement Embodied Carbon Tool) is a UK-developed spreadsheet tool that estimates both embodied and operational emissions for asphalt pavements by considering energy, water usage, materials, and other data, using emission factors for various stages of the asphalt lifecycle and formulas based on British design codes and standards (Huang *et al.*, 2013; Reeves *et al.*, 2020; Spray *et al.*, 2014). |
| Athena Pavement LCA tool (Athena Sustainable Materials Institute, 2024) | North America | Spreadsheet | Athena Sustainable Materials Institute database and the US Life Cycle Inventory Database | Cradle-to-grave | Tool based on material quantities | Material quantities | The Athena Pavement LCA tool is a software package that provides LCA results for materials manufacturing, roadway construction, and maintenance in Canada and the USA, supporting custom and pre-designed roadway configurations, including a comprehensive database of equipment and materials, and allowing users to specify unique pavement systems, input operating energy, and apply pavement vehicle interaction algorithms to refine LCA results, facilitating quick comparisons of multiple design options over various lifespans. |
| Atkins' Carbon Critical Knowledgebase (Atkins, 2024) | Worldwide | Web-based tool | Includes an extensive library of version-controlled operational and embodied GHG factor information developed by Atkins | Cradle-to-construction | Tool based on the contract | Material quantities and equipment (equipment details and human hours) | Atkins' Carbon Critical Knowledgebase, derived from ATKINS' project expertise, features a comprehensive list of structural elements and pre-configured combinations for civil infrastructure projects, allowing users to add new elements with specific details, input quantities in various units and expected lifespans, and generate reports based on construction contract bills of quantities, though the tool's empirical data and uncertainty levels remain largely confidential (Navaei *et al.* 2022; Prokofiev, 2014) |
| 'ERIC' Carbon Planning Tool (Environment Agency, 2024) | UK | Spreadsheet | Bath Inventory of Carbon and Energy, AggRegain Carbon Dioxide ($CO_2$) Emissions, British Department for Environment, Food and Rural Affairs (Defra), British Cement Association | Cradle-to-grave | Tool based on material quantities | Material quantities | The 'ERIC' Carbon Planning Tool converts the volume of structural elements to material quantities based on British standards, allows users to calculate material quantities to reduce uncertainty, provides options for material transport modes and route details, includes an optimisation feature requiring comprehensive data input, and offers alternative material options that users must verify for compliance with construction or design standards. |

| Name of tool | Location | Entry format | Background databases | Life cycle stages | Type of tool | Inputs | Description and characteristics |
|---|---|---|---|---|---|---|---|
| FHWA Infrastructure Carbon Estimator (Federal Highway Administration, 2014) | USA | Spreadsheet | Life-cycle Environmental Inventory of Passenger Transportation in the United States (Chester, 2008) Office of Highway Statistics Series, Highway Statistics 2012 (Office of Highway Policy Information, 2012) | Cradle-to-grave | Tool based on the final design | Final design | The FHWA Infrastructure Carbon Estimator simplifies emissions estimation using roadway distance as its primary input, which introduces high uncertainty due to generalisations and assumptions, requiring users to input general project information, construction and maintenance details, and construction delay, while offering mitigation strategies and adhering to US design standards, making it suitable for planning when detailed project knowledge is lacking (Raeisi et al., 2021). |
| National Highways Carbon Tool (National Highways, 2022) | UK | Spreadsheet | Similar with Carbon Planning Tool's database by the Environment Agency (UK) | Cradle-to-construction | Tool based on the contract | Material quantities and equipment (equipment details and human hours) | The National Highways Carbon Tool, similar to the ERIC Carbon Planning Tool, relies on construction contract data and processes data chronologically, includes a workflow format for calculating process durations without assessing workflow impact on emissions, requires input of quantities and transport distances for various materials and items, adheres to British standards with editable emission indicators, and provides guidance for each process (Giesekam et al., 2022; Reeves et al., 2020). |
| Klimatkalkyl Tool (Geokalkyl Tool for optimisation) (Swedish Transport Administration, 2018) | Sweden | 3D GIS Tool using Spreadsheet | Swedish Transport Administration's measurement database for environmental impact (TMO) | Cradle-to-grave | Tool based on the final design | Final design | The Klimatkalkyl Tool is an ArcGIS-based tool for calculating emissions in large-scale infrastructure projects in Sweden, using geometric inputs to create custom zoning and building types focused on environmental impacts, and offering a 3D representation feature for critical decision making (Strömberg et al., 2020) |
| Transportation LCA (Chester and Horvath, 2012a; 2012b) | USA | Web-based tool | transportation LCA database (tLCAdb) by Mikhail Chester, Arpad Horvath, and colleagues. | Cradle-to-grave | Tool based on the final design | Final design | The Transportation LCA Tool calculates emissions per distance travelled for various transport modes based on LCA theory, but its estimates have high uncertainty when generalised beyond the specific projects/studies it draws from. |
| VICE 2.0: Vehicle and Infrastructure Cash-Flow Evaluation Model (National Renewable Energy Laboratory, 2015) | USA | Spreadsheet | Database of State Incentives for Renewables and Efficiency (DSIRE) and data by original equipment manufacturers | Cradle-to-grave | Not including civil infrastructure | Equipment (equipment details and human hours) | VICE 2.0, developed by the National Renewable Energy Laboratory (NREL), is an empirical tool that collects data on alternative-fuelled and advanced vehicles, integrating it with EPA fuel economy data to estimate emissions for various transportation modes. |

(the asphalt level) based on the AASHTO standards. Only projects that had all the required data were taken into consideration.

For the purposes of designing the aggregate-surface roads, the AASHTO guidelines (2024) were used for all the roads presented here. Based on the design methodology, the inputs of the design process are the following (Nikolaides, 2015).

### 3.1 Traffic

Traffic data estimate Equivalent Single Axle Load (ESAL) units for road design, following AASHTO guidelines. Traffic is evenly distributed throughout the year, with seasonal distribution proportional to season lengths. Figure 1 shows maximum, minimum, and average cumulative traffic of the projects used in this research over 30 years for comparison, despite real projects having design lives of 25–40 years.

### 3.2 Climate

The year is divided into four seasons, according to the ground's moisture condition. These are

1. 'Winter', roadbed soil is frozen;
2. 'Spring/Thaw', roadbed soil is saturated;
3. 'Spring/Fall', roadbed soil is wet;
4. 'Summer', roadbed soil is dry.

The data for each project were collected by the Climate data for cities worldwide (2024).

### 3.3 Seasonal roadbed soil resilient moduli

To define the road strain, seasonal moduli for roadbed material were identified through lab tests in geotechnical reports, using samples with corresponding moisture, temperature, and stress conditions. The California bearing ratio (CBR) values were provided, and the resilient modulus ($M_R$) was estimated (Equation 1):

1. $M_R = 17.6 \cdot (CBR)^{0.64}$

### 3.4 Base/sub-base resilient moduli

For the resilient modulus of the base layer material, a constant value will be used. To calculate the modulus of elasticity of the base layer, either the equation to convert CBR values to $E_{BS}$ was used (Equation 1), or alternatively an equation that defines the modulus of elasticity in proportion to the layer thickness ($h_{bs}$) was used (Equation 2):

2. $E_{BS} = k \cdot M_R$

where $k = 0.2 \cdot (h_{bs})^{0.45}$

$2 < k < 4$

In all the cases, a base layer that corresponds to $E_{BS}$ = 30 000 psi was selected.

### 3.5 Design serviceability loss, allowable rutting, and aggregate loss

Design serviceability loss, allowable rutting, and aggregate loss were kept constant for all projects for comparison and not included in the analysis. According to AASHTO guidelines, road serviceability is measured by the present serviceability index (PSI), ranging from 0 (impossible to travel) to 5 (perfect condition). For aggregate-surfaced roads, the design serviceability loss is 3.0, meaning if the initial serviceability is 4.0, the terminal serviceability is 1.0. Rut depth failure refers to serious deformation of the pavement structure. Allowable rut depth for aggregate-surfaced roads depends on average daily traffic, typically ranging from 1.0 to 2.0 in. For design, an allowable rutting depth of 5.08 cm (2 in) was considered. Aggregate loss due to traffic and erosion is another concern. Estimating the total thickness of aggregate lost during the design period and the minimum required thickness is necessary. A conservative approach estimated a total aggregate loss (GL) of 0.35 cm (0.14 in).

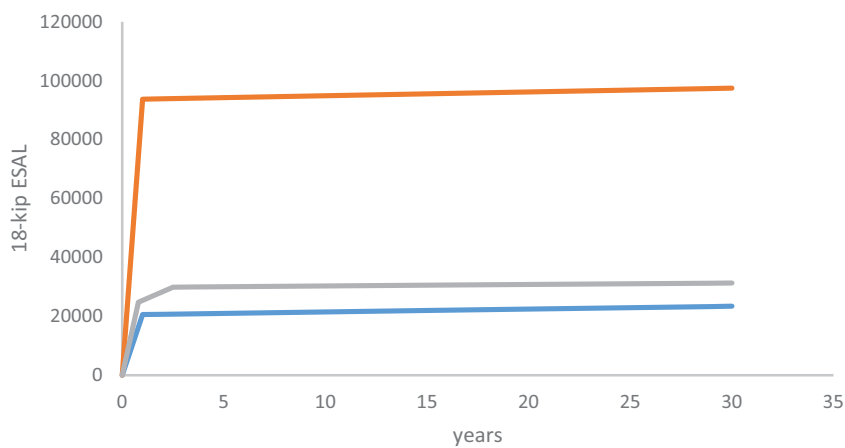

**Figure 1.** Maximum, minimum, and average cumulative traffic for 30 years period

### 3.6 Other and new design parameters

Other parameters are identified inductively by analysing the data of road projects. For the development of the appropriate parameter set, it was required to have access to the topographic data of areas, including the road layouts and to combine them with real data coming from the sites. Based on these data and using the Autodesk Civil 3D software, the material of every road was estimated (see the Supplementary Video file for the process followed). Autodesk Civil 3D was used for BIM-3D representation and to support carbon digital twinning, since is widely used in engineering for design.

The topographic data were used to create models of the existing surface. Civil 3D provided the proper tools for designing and analysing a new road. At first, the surfaces were created based on the data provided by the topographic data and the sites. Civil 3D created a new surface and automatically estimated the affected area and volume. The author had to correct the new surfaces, for example, the contours of the new surfaces should be in parallel with the route of the bulldozer or the grader during the earthworks.

The materials needed for constructing a road were estimated by comparing the existing surface with the new one. The output of the software in Figure 2 is the area that needs to be excavated (red colour in the road layout in Figure 2 online version) and the area that needs to be filled (green colour in the layout in Figure 2 online version).

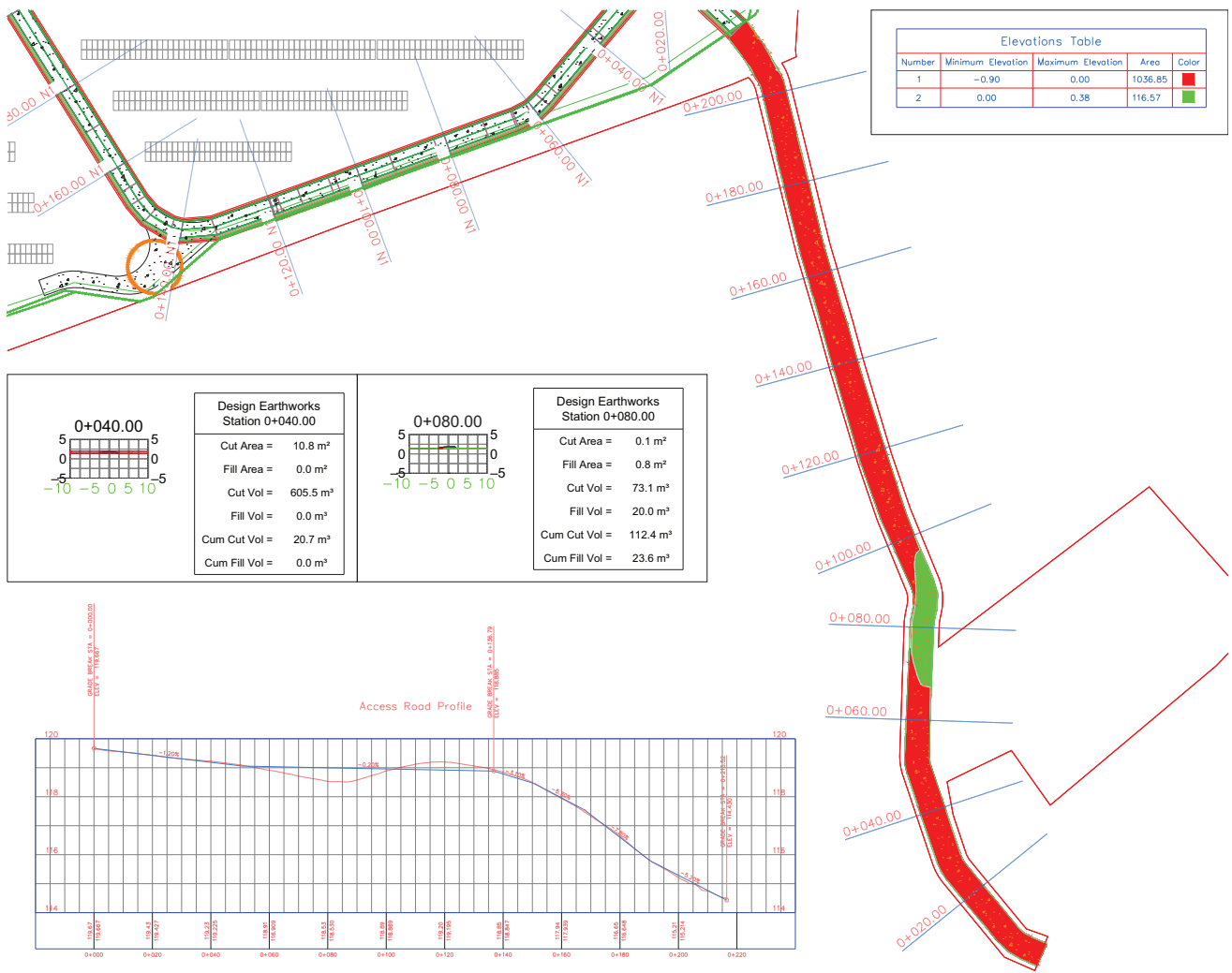

**Figure 2.** Civil 3D tool for road design and quantities

Regarding the design of a new road, using the topographical and geotechnical data, surfaces were created with different colours based on parameters, such as the CBR value, the level of the sub-phase, or any parameter of interest. This approach enabled the determination of optimal road alignment areas, thereby minimizing material usage by optimizing the parameter set's values.

In addition, Civil 3D provides the flow paths (see Figure 3 in the blue lines online version). The Geodesic data collected were combined with flood risk analysis data, such as the 'Check the long term flood risk for an area in England's maps' (UK GOV, 2024), to estimate the quantities of materials needed to build a drainage system to protect the road infrastructure from water. The drainage systems were designed with the Rational (Orthological) Method (Koutsogiannis, 2011; Papamichail, 2004). In Figure 3, someone can see how much earth material is required for levelling the existing surface to the new surface where the road and other infrastructure systems will be built. The material used (e.g. if concrete will be used in the ditches, riprap at the end of the flooding path etc.) is based on the hydraulic/drainage design.

Although Civil 3D provides data for road design, drainage, and earthworks, it does not offer final designs. It gives slopes for drainage systems but does not predict hydraulic jumps or design road assemblies based on traffic, climate, and so on. These parameters were considered by the author during the design process. Geosynthetic materials such as geogrid were not considered in this research, though they were used in the actual project to reduce aggregate use.

The design outputs were converted to emissions using ERIC's Carbon Planning Tool, focusing on embodied emissions. ERIC was chosen over other LCA tools for several reasons. Atkins'

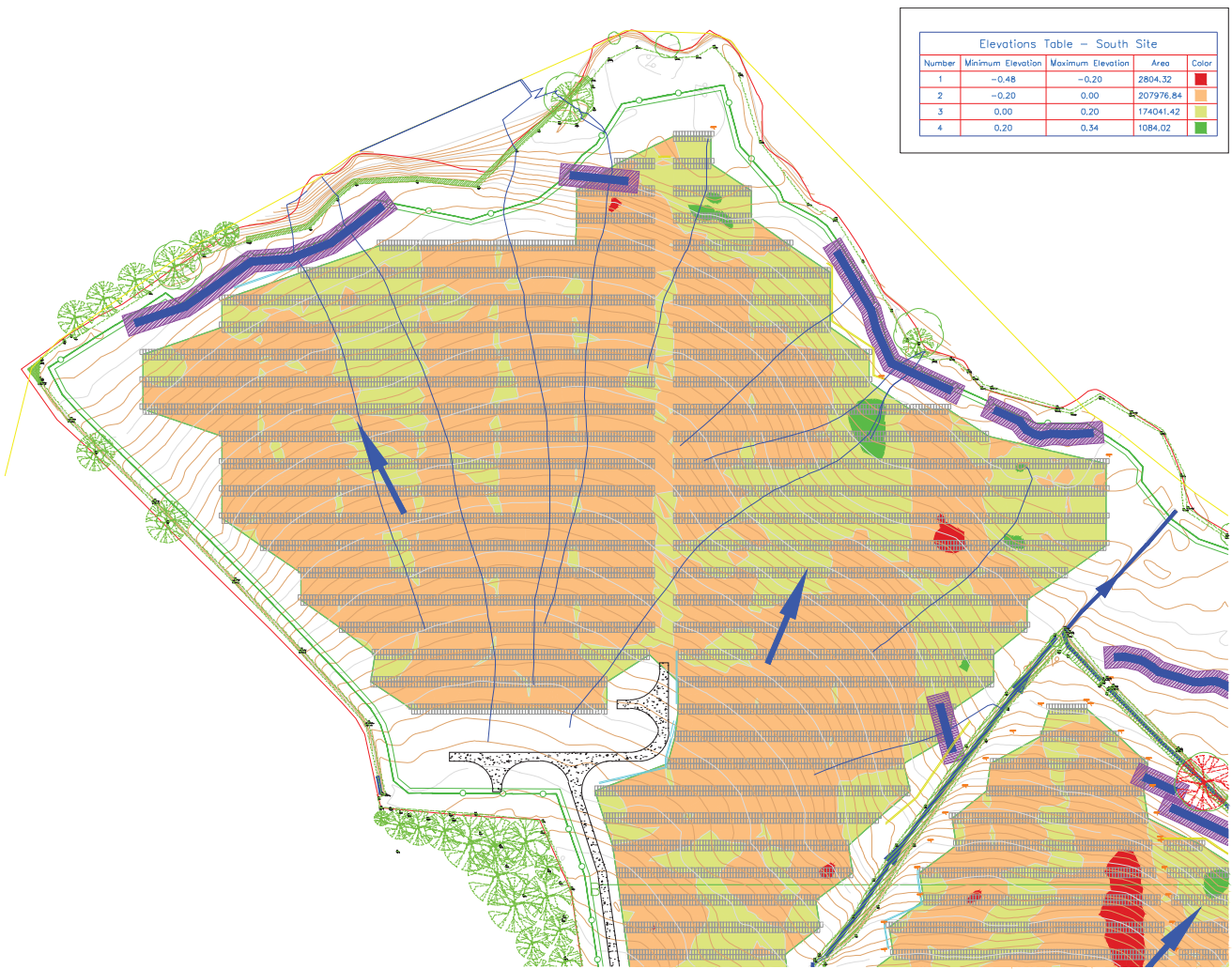

Figure 3. Earthworks and drainage systems

Carbon Critical Knowledgebase is not publicly accessible and requires adjustments for road design. Tools such as VICE 2.0, and Transportation LCA lack a top-down whole-life carbon assessment, crucial for understanding a project's full environmental impact. Athena Pavement LCA tool and FHWA Infrastructure Carbon Estimator are based on North American standards, unsuitable for UK and European roads. The Klimatkalkyl Tool, developed in Sweden, requires major adjustments.

ERIC, asPECT, and National Highways Carbon Tool are the most appropriate, with ERIC being preferred for its flexibility and detailed analysis. asPECT, designed for asphalt pavement, provides a detailed analysis of embodied carbon dioxide emissions but requires specific data that was not easily available for the 200 projects in this research. ERIC is better suited for estimating embodied carbon dioxide emissions of UK aggregate roads for several reasons. It provides a mechanism for assessing carbon over the whole life of built assets, crucial for understanding long-term environmental impacts. ERIC allows for top-down whole-life carbon assessment and options during project appraisal, informing the solution selection process efficiently. Unlike the National Highways Carbon Tool, ERIC enables deep analysis of inputs, separating embodied from operational emissions. It offers different levels of calculation, making it user-friendly and adaptable to various project scopes. ERIC also supports setting supplier carbon targets and promotes low-carbon solutions in construction works. Mentioned in government publications, ERIC aligns with national strategies and policies, adding to its credibility.

## 4. Results and analysis

The analysis was done with SPSS software (Field, 2009). Generally, the embodied emissions of roads were $880 \pm 240$ tCO$_2$eq per km for a single lane (3.5–4.0 m), estimated with a simple statistical analysis and the range of the values was a result of different parameters presented.

The effect of any parameter that emerges either deductively through literature review or inductively by analysing road designs were by statistical analysis (e.g. Pearson correlation, $t$ tests, Analysis of variance (ANOVA), Principal component analysis (PCA), linear regression models etc.) of completed projects (Field, 2009). ANOVA analysis processes was used for non-quantitative data, such as the soil types and surface 3D characteristics. The aforementioned analyses tools assume a linear relationship. To challenge the linearity that LCA assumes, the author examined parameters that affect the design using these tools and found that in some cases, no linear effect was detected. Constructability adjustments of the designed road were not considered during this analysis (e.g. the road level is rounded so that site engineers do not have to measure decimals), since they cannot be captured with these tools.

First, linear regression analyses and principal component analyses were run, to examine if there is a correlation between the parameters (variance inflation factor >10) (Field, 2009). The topographical and geometrical parameters of the road, such as the length, the slopes (on average), the number of intersections, the cross slopes (on average), the turning radius were correlated to each other. The width of the road was not correlated with the other geometrical parameters. All the parameters had a significantly positive correlation with the area of the road, meaning that the highest the value of the previous topographic parameter, the bigger the area of the road. So, the area of the road was decided to be used as a parameter instead of them. There was a significantly positive correlation between the area of the road and emissions generated. In addition, a negative correlation was revealed between the changes on the slope due to safety reasons, such as a reduction of the maximum slope to 8%–12% for heavy vehicles, and the emissions generated. Since the operation emissions were not taken into consideration in this study, the emissions for excavating works for reduction of the slope of the road due to safety reasons were not considered, but as someone can understand the embodied emissions were reduced. The statistical analysis showed no effect of the landscape inconsistency (this was tested both through Pearson correlation and ANOVA after grouping the slopes with different ways), meaning that changes on slopes have no effect on the embodied emissions. It worth noted that roads with landscape inconsistency produced more embodied emissions on average, but this was not captured with the aforementioned analyses.

The widths of the roads designed were either 3.5 m or 4.0 m. Since the width affects the area of the road should not be studied alone. Just for comparison reasons, it was explored whether 3.5 m and 4.0 m width produce differently emissions. Since there are only two groups −3.5 width against 4.0 m width independent $t$ tests were performed on the data. The results of the statistical analysis demonstrated that estimation of emissions did differ between the groups, the 4.0 m width roads produced more emissions than a 3.5 m width road, as expected.

After the road sections of each project were designed based on AASHTO, the level of the road layout was adjusted based on the flooding level. The level increased in areas where the road should be protected, and low water crossings were built in areas where the water should cross. We grouped the results based on the qualitative analysis the flooding maps of the UK do: low, medium, and high probability of flooding due to an area's long-term risk from (*1*) rivers and the sea, (*2*) surface water, (*3*) reservoirs, and (*4*) groundwater. One-way ANOVAs with flood groups as the independent variable and overall embodied emissions per km as the dependent one was performed on the data. The statistical analysis showed no effect of the flood risk, although the flood risk affects the design, the level, and the layout of the road.

Regarding the soil type, the Unified soil classification system was used to group the results in the 15 groups: GW, GP, GM, GC, SW, SP, SM, SC, ML, CL, OL, MH, CH, OH, and PT. The first letter means: G: gravel, S: sand, M: silt, C: clay, and O: organic, and the second P: poorly graded, W: well graded, H: high liquid limit, and L: low liquid limit. One-way ANOVAs showed that there was a main effect of the soil type ($p < 0.001$), indicating that embodied emissions per km generated were differentiated to a significant degree as a function of the soil type. Post-hoc comparisons with Bonferroni correction showed that the effect stemmed from the significant differences between the groups having produce the lowest against the groups that having produced the highest amount of emissions per km, that is well-graded gravel produced significantly lower emissions compared with fat clay ($p < 0.001$) and lean clay and ($p = 0.012$) the same applied for poorly graded gravel ($p < 0.001$ and $p = 0.006$, respectively). No other comparisons reached a significant level. The CBR of the soil was not correlated with the embodied emissions per km, although it was used for designing the road. The CBR is used for the initial design of the typical section of the road and then this section changes based on the other design parameters.

Since the design is a sequence of steps, it is challenging to capture its parameters with the linear assumptions, the LCA tools use.

The approach described above employs a variety of statistical tools to analyse the impact of different parameters on the embodied emissions of roads. The use of linear regression, ANOVA, and PCA is common in such analyses to identify correlations and the effects of various factors. The findings suggest that the area of the road is a significant parameter, positively correlated with emissions, which aligns with the expectation that larger areas would require more materials and energy, thus leading to higher emissions. The negative correlation between the reduction of slopes for safety reasons and emissions is also logical, as less material would be moved in such cases. The findings also highlights the complexity of road design and the difficulty of capturing all variables in a linear model, which is a valid consideration in LCA. The Unified Soil Classification System results indicate significant differences in emissions based on soil type, which is a valuable insight for planning and designing roads with environmental considerations in mind.

## 5. Conclusions

There is a need for more innovative and adaptive approaches that could optimise the performance, efficiency, and sustainability of road infrastructure projects. None of the transport LCA tools estimate the material quantities. The material quantities are an input for these tools coming from the design process. No tool exists that estimates the material quantities as part of the engineering design process. In addition, according to some recent publications (PIARC, 2022), the state-of-the-art in road infrastructure design is still based on conventional standards and guidelines that may not adequately address the requirements of a rapidly changing world in the field of road transport systems, such as new technologies, changes in mobility modes, and availability of multiple diffused data sources.

The study conducted a thorough analysis of various topographical and geometrical parameters of road design to understand their correlation with emissions generated. The findings revealed significant positive correlations between the area of the road and emissions, while the width of the road did not correlate with other geometrical parameters. Interestingly, changes in slope for safety reasons showed a negative correlation with emissions, indicating a reduction in embodied emissions due to slope adjustments. The analysis also highlighted that roads with landscape inconsistency produced more embodied emissions on average, although this was not statistically significant. The comparison of road widths demonstrated that wider roads (4.0 m) generated more emissions than narrower roads (3.5 m). Flood risk, despite influencing road design, did not show a significant effect on embodied emissions. However, soil type had a notable impact, with well-graded gravel producing significantly lower emissions compared with fat clay and lean clay. The CBR of the soil, although used in road design, did not correlate with embodied emissions. Overall, the study underscores the importance of considering various design parameters and their interactions to minimise emissions in road construction. Future research should explore non-linear relationships and other factors that may influence emissions, providing a more comprehensive understanding of sustainable road design practices.

**How can you contribute?**

To discuss this paper, please email up to 500 words to the editor at support@emerald.com. Your contribution will be forwarded to the author(s) for a reply and, if considered appropriate by the editorial board, it will be published as discussion in a future issue of the journal.

*Proceedings* journals rely entirely on contributions from the civil engineering profession (and allied disciplines). Information about how to submit your paper online is available at www.icevirtuallibrary.com/page/authors, where you will also find detailed author guidelines.